\documentclass[aps,twocolumn,groupedaddress]{revtex4}
\usepackage{bm}
\usepackage{graphicx}

\begin{document}
\bibliographystyle{revtex}

\title{Relative influences of disorder and of frustration on the glassy
dynamics in magnetic systems}

\author{F. Ladieu$^{1,*}$, F. Bert$^2$, V. Dupuis$^3$, E. Vincent$^1$, J. Hammann$^1$}

\affiliation{$^1$ DSM/DRECAM/SPEC C.E.A. Saclay, 91191 Gif/Yvette, France; $^2$ Laboratoire de 
Physique des 
Solides,
Universit\'e d'Orsay Bat. 510, 91400 Orsay, France; $^3$ L.M.D.H., U.P.M.C. CNRS, 4 place 
Jussieu, 
75252 Paris 
C\'edex 05, 
France.}

\date{\today}

\begin{abstract}
The magnetisation relaxations of three different types of geometrically frustrated magnetic 
systems 
have been studied with the same experimental procedures as previously used in spin glasses. 
The materials investigated are Y$_2$Mo$_2$O$_7$ (pyrochlore system), 
SrCr$_{8.6}$Ga$_{3.4}$O$_{19}$
(piled pairs of kagome layers) and (H$_3$O)Fe$_3$(SO$_4$)$_2$(OH)$_6$ (jarosite compound). 
Despite a very small amount of disorder, all the samples exhibit many characteristic features
of spin glass dynamics below a 
freezing temperature $T_g$, much smaller than their Curie-Weiss temperature $\theta$. The aging 
properties 
of their thermoremanent magnetisation can be well accounted for by the same scaling law as in 
spin glasses and the values of the scaling exponents are very close. The effects of temperature 
variations during aging have been specifically investigated. In the pyrochlore and the bi-kagome
compounds, a decrease of temperature after some waiting 
period at a certain temperature $T_p$ re-initialises aging and the evolution at the new 
temperature is the 
same as if the system
were just quenched from above $T_g$. However as the temperature is heated back to $T_p$, the 
sample 
recovers the state it had previously reached at that temperature. These features are known in 
spin glasses as
rejuvenation and memory effects. They are clear signatures of the spin glass dynamics. In the 
kagome
compound, there is also some rejuvenation and memory, but much larger temperature changes are 
needed 
to observe the effects. In that sense, the behavior of this compound is quantitatively different 
from that 
of spin glasses.

\end{abstract}


\maketitle

\begin{figure}
\includegraphics[height=18.5cm, width=8.5cm]{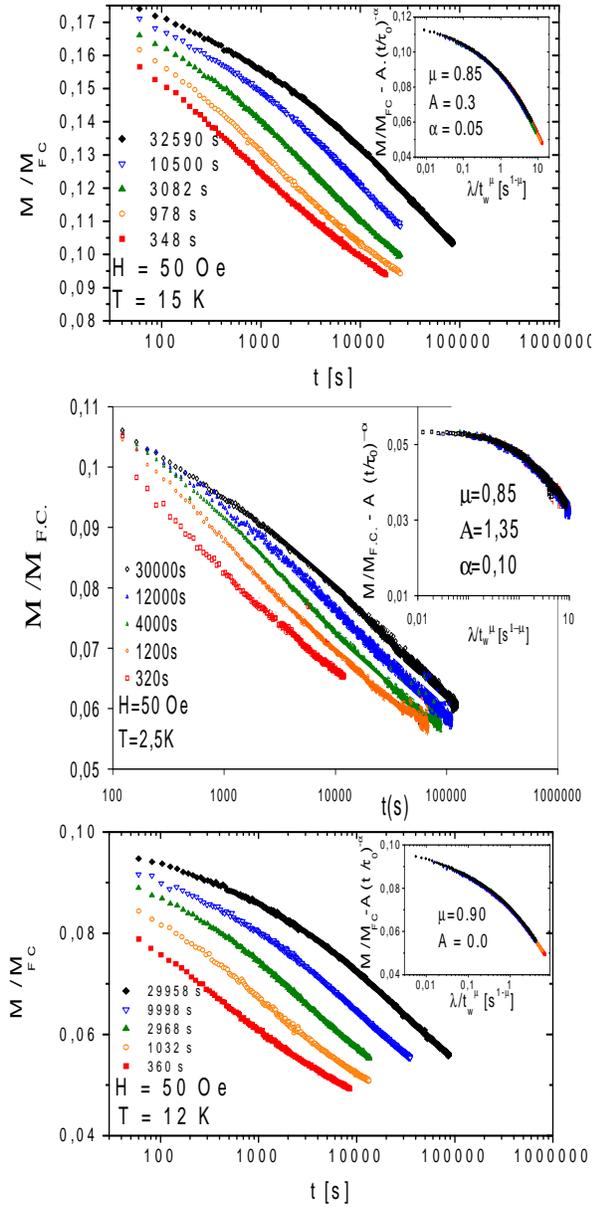}
\caption{Thermo Remanent Magnetization experiments carried out at $T=0.7 \times T_g$ (see text). 
>From 
top to bottom: 
YMO, 
SCGO, kagome samples. The longer the waiting time $t_w$, the slower the 
relaxation when the magnetic field is switched off. Just as in standard spin glasses, the aging 
magnetization 
$M(t,t_w)-A(\tau_0/t)^{\alpha}$ can be scaled (see the insets) as a function of 
$\lambda/t_{w}^{\mu}$, where 
$\lambda = 
t_w[(1+t/t_w)^{1-\mu}-1]/[1-\mu]$, i.e. $\lambda/t_w^\mu \simeq t/t_w^\mu$ for $t \lesssim t_w$. 
The 
best fit gives  
parameters very close to those found in standard spin glasses. For 
YMO $\mu=0.85$, $\alpha=0.05$, $A=0.3$; for SCGO $\mu = 0.85$, $\alpha=0.10$, $A=1.35$; for the 
kagome sample 
$\mu=0.90$ 
and 
$A \simeq 0.0$.}
\label{F1}
\end{figure}

It was usually thought that both disorder and frustration were necessary to obtain spin glass 
dynamics (e.g.\cite{Lundgren}, \cite{VincentCM1996}, \cite{Bouchaud}). However, several 
frustrated 
but non-disordered antiferromagnets, rather than condensing into a spin liquid state as 
theoretically 
expected, were found to enter a glassy state below a 
well 
defined freezing temperature. As in usual Spin Glasses, below 
this temperature T$_g$, their dynamic response to a small magnetic field becomes very slow and 
depends 
on the time $t_w$
spent below T$_g$ (`aging', see e.g.\cite{Lundgren}, \cite{VincentCM1996}, \cite{Bouchaud}). The 
question thus arises of whether these glassy materials are really similar to standard spin 
glasses. 
We have addressed this question by investigating three examples of such antiferromagnets using 
the same 
experimental 
procedures as previously used in the detailed study of the dynamical properties of spin glasses.   
The three examples correspond each to a different topology of the frustrated spin network. 
The first one is Y$_2$Mo$_2$O$_7$ (`YMO')  where the spins ($S=3/2$) are on a 
three-dimensionnal 
($3d$) pyrochlore lattice \cite{Gingras}, \cite{Booth}. The second one is 
SrCr$_{8.6}$Ga$_{3.4}$O$_{19}$, 
(`SCGO') where the spins are located on bi-layers of $2d$ kagome lattices, each bi-layer being 
well 
separated
from the others \cite{Ramirez}, \cite{Keren}. The third one is a 
Jarosite compound (H$_3$O)Fe$_3$(SO$_4$)$_2$(OH)$_6$ where the spins ($S=5/2$) are located on 
genuine 
$2d$ kagome 
layers, 
with small inter layer couplings \cite{WillsPRBR2000}.

\begin{figure}
\includegraphics[height=18.5cm, width=8.5cm]{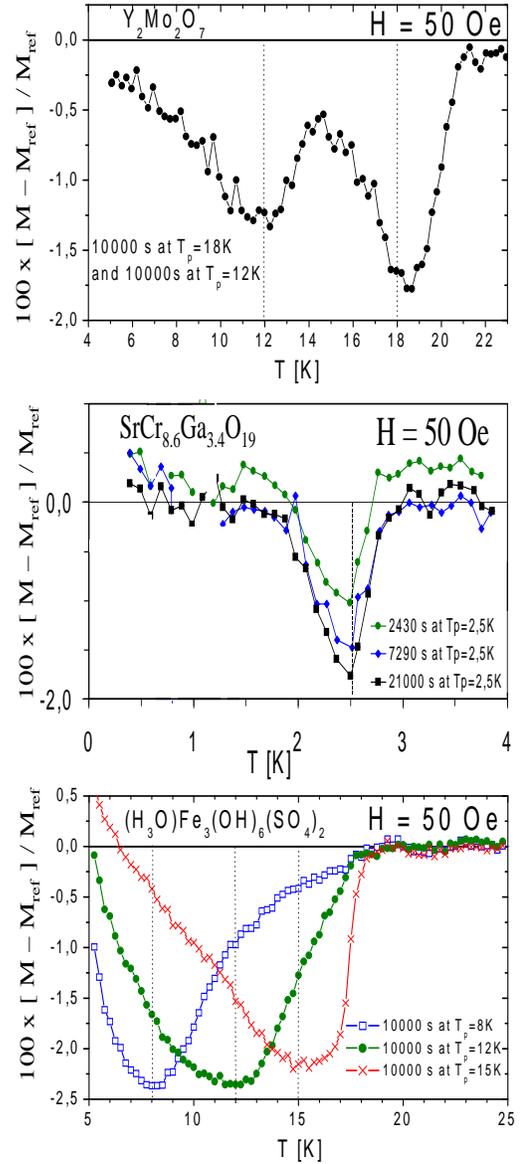}
\caption{ Memory dip experiment (see text): $M_{ref}$ is measured in a standard ZFC experiment, 
while $M(T)$ corresponds to the experiment where the cooling was 
interrupted for a time $t_w$ at a temperature $T_p$ (marked by the dotted lines). 
>From top to bottom: YMO, SCGO, kagome samples. The width of the 
memory dip is quite 
small in YMO and in SCGO (just as in standard spin glasses). It turns out to be much larger in 
the kagome sample.}
\label{F2}
\end{figure}

At high temperature, all three compounds exhibit a Curie Weiss behaviour with a 
large negative value for the Curie-Weiss temperature $\theta$, indicating strong 
antiferromagnetic 
interactions between spins. The values of $\theta$ are much larger than the observed freezing 
temperatures $T_g$. 
The ratio $\vert \theta \vert /T_g$ is of order $10$ for the (3d) YMO sample \cite{Gingras}, 
about 
$140$ 
for SCGO \cite{Ramirez}, and about $60$ for the (2d) kagome sample \cite{WillsPRB2000}. This 
reveals 
that 
frustration plays a key role in the low-$T$ physics.

Below $T_g$ a clear separation between Zero Field Cooled (ZFC) and Field Cooled (FC) 
magnetisations 
is found. This 
is a 
standard indication suggesting that the low-temperature phase is a spin glass phase. The spin 
glass 
transition is 
further 
confirmed by other 
measurements: (i) 
in the kagome sample an analysis of the critical dynamics was carried out over 10 decades in 
frequency (including 
M\"ossbauer 
experiments) and it yielded the critical exponents 
usually found in standard spin glasses \cite{DupuisPhD2002}; (ii) in 
YMO \cite{Gingras} it was shown that the nonlinear susceptibility diverges at $T_g$, as expected 
for 
spin glasses. 
Besides, 
both in YMO \cite{Booth} and in the kagome sample \cite{WillsPRB2001}, neutron diffraction 
experiments 
do not 
detect any 
long range ordering. 

The amount of disorder is very small in all these materials. In YMO the main source of disorder 
might be a few 
percents 
of $Mo-Mo$ bonds whose lengths are slightly different from the standard one \cite{Booth}. In SCGO 
an obvious 
source of 
disorder could arise from the non stoechiometry of the compound which leads to a $Cr$  
coverage 
of the magnetic sublattice of $x=8.6/9 = 95\%$, i.e. to an amount of disorder of $5\%$. However, 
since it was 
shown that 
$T_g$ 
is an increasing function of $x$ \cite{Ramirez}, one expects a spin glass phase even in the 
stoechiometric 
compound. 
Thus, the imperfect coverage of the magnetic sublattice does not seem to trigger the spin glass 
phase. Last, 
in the Kagome sample \cite{WillsPRB2000}, the coverage of the $Fe$ network is higher than $97\%$, 
leading, as in 
the 
two other compounds, to an upper bound of a few percents of disorder. 

To further investigate the spin glass phase in these samples, we first performed Thermo Remanent 
Magnetization 
(TRM) experiments. In the measurements described here, the sample is quenched from above $T_g$ to 
$T_m=0.7\times 
T_g$ 
in a magnetic field of $H=50\ $Oe. After a certain waiting time $t_w$ at $T_m$, the field is 
switched off and the 
magnetization $M$ is recorded as a function of time $t$. The results are shown in Fig.1. They 
look 
very similar to 
those of standard spin glasses, displaying a very slow $t_w$ dependent relaxation. The larger 
$t_w$, 
the slower the 
relaxation, and the larger the remanent magnetization at a given time $t$. This indicates that 
more 
and more 
spin correlations develop as time $t_w$ becomes larger.
\cite{Lundgren,VincentCM1996,Bouchaud}, \cite{Alba}. 

As can be seen in Fig.1, the whole set of TRM experiments yields results very similar to those of 
spin glasses. Moreover, applying the standard scaling procedure used in spin glasses, 
(see, e.g. \cite{VincentCM1996}) 
allows a perfect scaling of  
all the 
$M(t,t_w)$ data on a unique curve. This is obtained by fitting the relaxation curves 
to an expression 
with two additive terms. The first one is a stationary ($t_w$ independent) power law relaxation 
$A(\tau_0 
/t)^{\alpha}$
which is predominant at very small times. The second term is 
a function 
$F(t,t_w)$ 
of $t$ and $t_w$. This function is found to scale as $t/t_w^{\mu}$ at short $t$ with $\mu$ being 
an 
exponent close 
to 1. 
In this short $t$ region, $t_w$ is approximately the age of the system i.e. the time spent at the 
measuring 
temperature.
At larger $t$, the real age must be taken as $t+t_w$, in other words, the system continues to age 
as 
one measures 
its 
relaxation. It can then be shown that the correct scaling variable should be rewritten as 
$\lambda/t_w^{\mu}$ where 
$\lambda = t_w[(1+t/t_w)^{1-\mu}-1]/[1-\mu]$. Note that one recovers $\lambda/t_w^\mu \simeq 
t/t_w^\mu$ for $t 
\lesssim 
t_w$, as expected. The insets of 
Fig.1 show the plots of the function F versus this reduced variable for the present samples. The 
scaling is shown 
to
apply quite well. Setting the microscopic attempt time to $\tau_0=10^{-12}$s, one finds fitting 
parameters very close 
to 
those of the standard spin glasses (see the caption of Fig.1), especially for the aging 
exponent 
$\mu$ which is slightly below $1$, indicating the usual `sub-aging' 
phenomenon. 

Beyond these standard TRM measurements which are performed at constant temperature (after the 
initial 
quench), 
experiments
involving temperature variations during aging below $T_g$ usually reveals very striking 
characteristic 
features in standard  
spin-glasses \cite{Mathieu}, \cite{DupuisJAP2002}. For example, it is now well established that 
aging at any $T_p<T_g$ has no apparent effect on the state at all 
other temperatures 
sufficiently different from $T_p$. This was found, for instance, in ZFC experiments in which 
modified cooling 
protocols
were applied \cite{Mathieu}, \cite{DupuisJAP2002}. While in the usual ZFC protocol the system is 
cooled (at a constant cooling rate) in zero field from above $T_g$ 
down to the lowest temperature, in the modified 
protocols, the cooling is 
interrupted at some temperature(s) $T_p$ and resumed after some waiting time $t_w$. In both 
cases, a small 
dc field is applied at the lowest temperature reached, and the magnetisation is recorded while 
the 
system is heated up at a constant rate. 
The ZFC magnetisation, as measured in the modified protocol clearly showed
marked
singularities (`dips') centered around the temperature(s) $T_p$. Far enough from $T_p$, it 
recovered 
the values of the usual 
(reference) ZFC magnetisation (\cite{Mathieu}, \cite{DupuisJAP2002}). This showed 
that aging at $T_p$ during $t_w$ did  
not affect the system at a temperature $T$ different from $T_p$. It also implied that the system 
kept the memory of aging at $T_p$
while it was at lower temperatures and was able to retrieve the aged state previously reached at 
$T_p$. These results proved  
that in spin glasses different spin correlations are building up at different temperatures and 
that 
the correlations built up at any 
temperature remain imprinted at lower temperatures. \textit{In these experiments, the width of 
the 
dips gives an idea of the `temperature selectivity' of the spin correlations}.
The results of such ZFC experiments are displayed in Fig.2 for the three considered compounds. In 
this figure, the 
differences between the modified ZFC and the reference ZFC data are plotted as a function of 
temperature. Clear 
singularities are observed around the various chosen values of $T_p$ exactly as in spin-glasses. 
The 
($3d$)YMO 
sample as 
well as the SCGO material present narrow dips characterised 
by a value 
$\delta T/T_g=0.17 \pm .03$ very close to the values found in standard spin 
glasses: for instance, $\delta T/T_g=0.16 \pm.03$ in the $CdCr_{1.7}In_{0.3}S_4$ Heisenberg spin 
glass for 
comparable $t_w$, while $\delta T/T_g=0.21 \pm.03$ in the $Fe_{0.5}Mn_{0.5}TiO_3$ Ising spin 
glass. 
This is in contrast with the results on the kagome ($2d$) compound 
where $\delta T/T_g = 0.39 \pm .05$ is remarkably larger. 

The `temperature selectivity' has been studied more quantitatively by measuring the effect at 
some 
temperature $T$ 
of 
aging at a slightly smaller temperature $T-\Delta T$. This experiment yields a TRM measurement 
with 
the following 
procedure : (i) after field cooling from above $T_g$ to $T$, the sample is kept for a short time 
$t_1=500$ s 
at constant $T$, (ii) it is then further cooled to $T-\Delta T$ and aged during $t_2=9000$s 
before 
being re-heated 
to $T$ 
where it stays for another short time $t_3=t_1$, (iii) last, the field is switched off and the 
TRM 
curve is 
recorded. 
>From these experiments, an effective time $t_2^{eff}$ can be defined such as to superimpose the 
measured TRM with a purely isothermal TRM recorded at temperature $T$ with waiting time 
$t_1+t_2^{eff}+t_3$. In other 
words, 
$t_2^{eff}$
is a measure of the effect at $T$ of aging at $T-\Delta T$ for a time $t_2$. 
The actual quantitative results for the presently studied materials are reported in Fig.3 and 
compared to 
those obtained in  Ref\cite{Bert} for the 
$CdCr_{1.7}In_{0.3}S_4$ Heisenberg spin glass and for the $Fe_{0.5}Mn_{0.5}TiO_3$ Ising spin 
glass.
The results for YMO and SCGO are clearly in agreement with those of spin glasses, the slopes of 
$t_2^{eff}/t_2$ versus $\Delta T$ for these two systems are indeed within the values delimited by 
the 
Heisenberg and Ising cases. As for the Fe-jarosite, the effective time $t_2^{eff}$ does not vary 
much with the change of temperature. The system ages almost as quickly at $T-\Delta T$ than at 
$T$, 
and builds up the same pattern of spin correlations at both temperatures.

The conclusion of this investigation is that, in spite of their very small amount of disorder, 
the 
two frustrated antiferromagnets, the YMO pyrochlore 
and the bi-kagome SCGO, behave exactly as the standard spin glasses. The Fe-jarosite kagome 
compound, 
though presenting many similarities in terms of the existence of a freezing temperature and the 
presence of aging 
effects,
has quite different dynamical properties as a function of temperature. It appears, indeed, in 
Fig.3 
that $t_2^{eff}$ remains very close to $t_2$ even for larger $\Delta T$'s, meaning that aging at 
$T-\Delta T$ 
contributes significantly to the age of the 
system at $T$. The behavior is not very temperature selective implying that the pattern of the 
spin correlations
is not very temperature dependent. The two-dimensional character of the jarosite compound may 
be at the origin of the observed differences with the two other materials. 
\cite{Ritchey}.

\begin{figure}
\includegraphics[height=10.5cm, width=8.5cm]{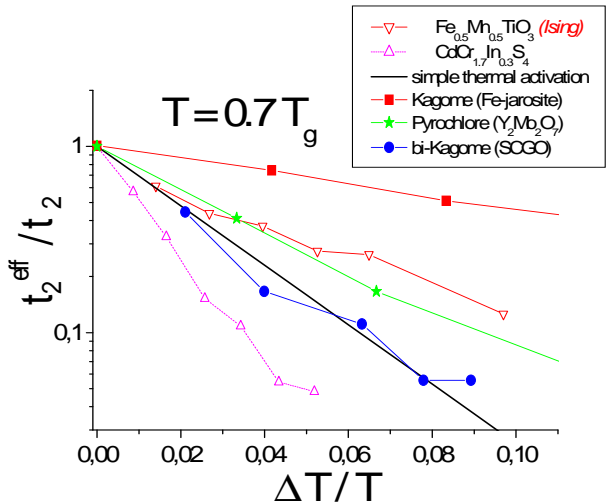}
\caption{$t_2^{eff}/t_2$ plotted versus $\Delta T/T$ (see text): the data for 
$Fe_{0.5}Mn_{0.5}TiO_3$ 
and 
$CdCr_{1.7}In_{0.3}S_4$ are plotted as examples of, respectively, an Ising spin glass and an 
Heisenberg spin 
glass. 
The 
steeper the slope in this plot, the more `temperature-selective' the aging (solid line: 
prediction 
assuming that 
aging 
occurs via thermal activation with an activation energy independent on $T$). In contrast to SCGO 
and 
YMO data which 
fall 
within the standard spin glass area, the kagome data are well above this region, with a very 
small 
slope.}
\label{F3}
\end{figure}

Many thanks to A.S. Wills (Edimburg University, United Kingdom) for giving us the Kagome sample, 
to 
N. Blanchard 
(LPS, 
Orsay, France) and to G. Collin (LLB, CEA Saclay) for preparing the SCGO sample previously 
studied 
by NMR by Ph. 
Mendels 
(Orsay University, France), and to J. Greedan (McMaster University, Canada) for 
giving us the YMO sample. The technical help of 
L. Le Pape is greatly acknowledged.

\vspace{7mm}

* ladieu@drecam.cea.fr


\begin{thebibliography}{99}

\bibitem{Lundgren} L. Lundgren, P. Svedllinh, P. Nordblad, O. Beckman, Phys. Rev. Lett. {\bf 51}, 
911 (1983).
\bibitem{VincentCM1996} E. Vincent, J. Hammann, M. Ocio, J.-P. Bouchaud, L.F. Cugliandolo, 
cond-mat/9607224.
\bibitem{Bouchaud} J.-P. Bouchaud, V. Dupuis, J. Hammann, E. Vincent, Phys. Rev. B {\bf 65},
024439 (2002).

\bibitem{Gingras} M.J.P. Gingras, C.V. Stager, N.P.Raju, B.D. Gaulin, J.E. Greedan, Phys. Rev. 
Lett., {\bf 78}, 
947, 
(1997).

\bibitem{Booth} C.H. Booth, J.S. Gardner, G.H. Gwei, R.H. Heffner, F. Bridges, M.Z. Subramanian, 
Phys. Rev. B, {\bf 
62}, 
R755 (2000); L. Bellier-Castella, M.J.P. Gingras, P.C.W. Holdsworth, R. Moessner, 
Cond-Mat/0006306.

\bibitem{Ramirez} A.P. Ramirez, G.P.Espinosa, A.S.Cooper, Phys. Rev. B, {\bf 45}, 2505, (1992).
\bibitem{Keren} A. Keren, Y.J. Uemura, G. Luke, P. Mendels, M. Mekata, T. Asano, Phys. Rev. 
Lett., 
{\bf 84}, 3450, 
(2000).


\bibitem{WillsPRBR2000} A.S. Wills, V. Dupuis, J. Hammann, R. Calemczuk, Phys. Rev.
B Rapid Comm. {\bf 62}, R9264 (2000).

\bibitem{WillsPRB2000} A.S. Wills, A. Harrison, C. Ritter, R.I. Smith, Phys. Rev. B, {\bf 61}, 
6156, 
(2000).

\bibitem{DupuisPhD2002} V. Dupuis, PhD Thesis, Paris XI Orsay, (2002).


\bibitem{WillsPRB2001} A.S. Wills, G.S. Oakley, D. Visser, J. Frunzke, A. Harrison, H.Andersen, 
Phys. Rev. B, {\bf 
64}, 
094436, (2001).

\bibitem{Alba} M. Alba, M. Ocio, J. Hammann, Eur. Phys. Lett., {\bf 2}, 45, (1986); E. Vincent, 
J. 
Hammann, M. 
Ocio, J.-P. 
Bouchaud, L.F. Cugliandolo, in \textit{Complex Behavior of Glassy Systems}, Lecture Notes in 
Physics 
Vol. 492 
(Springer-Verlag, Berlin, 1997), p.184 and references therein.

\bibitem{Mathieu} R. Mathieu, P. J\"onsson, D.N.H. Narn, P. Nordblad, Phys. Rev. B {\bf 63}, 
092401 
(2001); T. 
Jonsson, K. 
Jonason, P. J\"onsson, P. Nordblad, Phys. Rev. B, {\bf 59}, 8770 (1999).
\bibitem{DupuisJAP2002} V. Dupuis, E. Vincent, J. Hammann, J.E. Greedan, A.S. Wills, Journal of 
Applied Physics, 
{\bf 91}, 
8384, (2002).


\bibitem{Bert} F. Bert, V. Dupuis, E. Vincent, J. Hammann, J.-P. Bouchaud, cond-mat/0305088.

\bibitem{Ritchey} I. Ritchey, P. Chandra, P. Coleman, Phys. Rev. B, {\bf 47}, 15342, (1993); P. 
Chandra \textit{et 
al.}, J. 
Phys. I, {\bf 3}, 591, (1993).


\end{thebibliography}
\end{document}